\documentclass[journal=jpclcd,manuscript=article,layout=twocolumn]{achemso}
\usepackage{amsmath}
\usepackage{amssymb}
\usepackage{bm}
\usepackage{braket}	
\usepackage{color}
\title{An exciton-coupled electron transfer process controlled by non-Markovian environments}

\author{Souichi Sakamoto}
\email{sakamoto@kuchem.kyoto-u.ac.jp}
\affiliation{Department of Chemistry, Graduate School of Science,
Kyoto University, Sakyoku, Kyoto 606-8502, Japan}
\author{Yoshitaka Tanimura}
\email{tanimura.yoshitaka.5w@kyoto-u.ac.jp}
\affiliation{Department of Chemistry, Graduate School of Science,
Kyoto University, Sakyoku, Kyoto 606-8502, Japan}

\date{\today}

\begin{document}

\begin{tocentry}
 \centering
 \includegraphics[clip, width = 5.0cm]{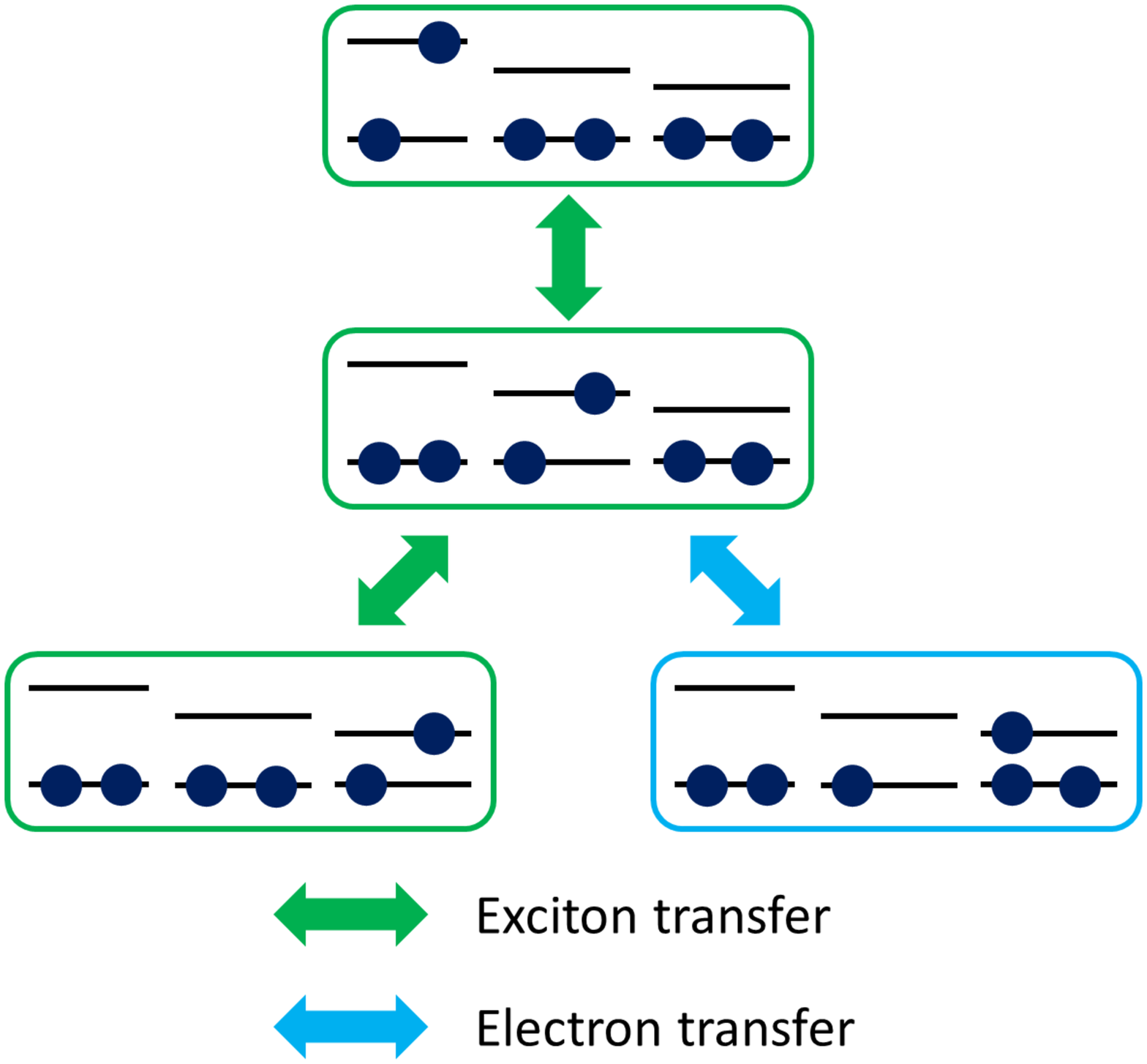}
\end{tocentry}

\begin{abstract}
We theoretically investigate an exciton-coupled electron transfer (XCET) process that is conversion of an exciton into a charge transfer state.
This conversion happens in an exciton transfer (XT) process, and the electron moves away in an electron transfer(ET) process in multiple environments (baths). This XCET process plays an essential role in the harvesting of solar energy in biological and photovoltaic materials. We develop a practical theoretical model to study the efficiency of XCET process that occurs either in consecutive or concerted processes under the influence of non-Markovian baths. The role of quantum coherence in the XT-ET system and the baths is investigated using reduced hierarchal equations of motion (HEOM). This model includes independent baths for each XT and ET state, in addition to a XCET bath for the conversion process. We found that, while quantum system-bath coherence is important in the XT and ET processes, coherence between the XT and ET processes must be suppressed in order to realize efficient irreversible XCET process through the weak off-diagonal interaction between the XT and ET bridge sites arises from a XCET bath. 
\end{abstract}

\maketitle	
Energy conversion of an exciton into a charge transfer state, referred to as ``an exciton-coupled electron transfer (XCET) process'', plays an essential role in many types of biological photosystems and functional molecular materials as the basic mechanism for utilizing solar energy. Examples include photosynthetic\cite{SchultenJCP2009,  SchultenJCP2011, SchultenJCP2012, Ishizaki2009, KramerFMO, SchultenFMO, Renger2005, Nakatsuji1998, Yan2017,Renger2012, Renger2015, Renger2017, Valkunas2017, IshizakiJCP15, Sumi96, Nov2011, Nov2015, Mukamel2013, Kramer17, Shi2012}and solar battery systems.\cite{Coropceanu2009,Gelinas2014, Thoss2015, Tamura2013, TamuraJPC2015, Prior2017} 

While these XT and ET processes themselves have been studied extensively, the mechanism of the irreversible XCET process, has not yet been established. This conversion mechanism is important not only as a fundamental process in chemistry and biochemistry but also for applications, such as those to organic photovoltaic materials used in solar energy harvesting devices. 
In this letter, we present a simple model {to analyze} the XT, ET, and XCET processes {for a variety of system involving photosynthetic systems and photovoltaic materials}. Using this XT-ET model, we investigate XT-ET conversion dynamics in non-Markovian environments (baths). In this study, special attention is paid to the role of the baths, because the irreversibility in the XCET process arises from the influence of the baths. We numerically determined the time evolution of the reduced density matrix elements by employing the hierarchy equations of motion (HEOM). With this method, we are able to properly treat the effect of the system-bath interaction on the XT-ET conversion dynamics. \cite{Tanimura88,Ishizaki05,Tanimura06,Tanimura14,Tanimura15,Arend2015,TMJPSJ1994,TanakaJPSJ09,TanakaJCP10,TanimruaJCP12}\\

\begin{figure}[tbp]
  \centering
  \includegraphics[width = 7.5cm]{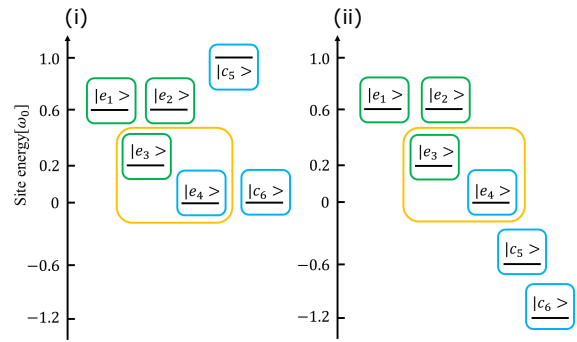}
\caption{{Schematic view of the (i) up-and-down and (ii) downhill models} described by the Hamiltonian given in eqs. (1) and (2). Here, $|e_i\rangle$ and $|c_j\rangle$ represent the XT and ET states at the site $j$. The green, blue and orange squares represent the XT bath of the antenna system, the ET bath of the reaction center, and the XCET bath for XT and ET bridge states, respectively.}
  \label{fig:model}
\end{figure}

Although the framework employed for the present model can be applied to  {a variety of system,} here we consider the particular case of a system with $N$ chlorophyll sites as a practical model to {learn from} the fundamental features of photosynthetic systems. In this system, each of the chlorophyll sites is characterized by the HOMO and LUMO levels, as illustrated in the {graphical} TOC. The XT and ET states of the $j$th site are represented by $|e_j\rangle$ and $|c_j\rangle$, respectively. {As illustrated in ref. \citenum{Arend2015}, we may include a grand and hole states in each site in addition to the XT and ET states, but they can be ignored in a case of photosynthetic system.}
The states from $j=1$ to $j=n_{XT}$ are regarded as the XT states, while the states from $j=n_{XT}$ to $N$ are regarded as the ET states. As illustrated in Fig. \ref{fig:model}, the site $j=n_{XT}$ involves both the XT and ET states (XCET states). Thus, the conversion from the excited states of the pigments in the antenna complex to those of the reaction center is realized at the sites $j=n_{XT}-1$ and $j=n_{XT}$. \\

The Hamiltonian of the system is\cite{Arend2015} 
\begin{eqnarray}
{{\cal \hat H}_S} &=& \sum_{i=1}^{n_{XT}} {{\rm \varepsilon _i^{(XT)}}|{e_i} \rangle  \langle {e_i}|} +\sum_{j=n_{XT} + 1}^{N} {{\rm \varepsilon _j^{(ET)}}|{c_j} \rangle  \langle {c_j}|} \nonumber \\ 
 &+& \sum_{i = 1}^{n_{XT}-1} ({J_{i, i+1}}|{e_i} \rangle  \langle {e_{i+1}}| + h.c.) \nonumber \\
&+& {t_e}(|{e_{n_{XT}}} \rangle  \langle {c_{n_{XT} + 1}}| + h.c.) \nonumber \\
&+& \sum_{j=n_{XT} + 1}^{N-1}{t_e}(|{c_{j}} \rangle\langle {c_{j+1}}| + h.c.),
\end{eqnarray}
where ${\rm \varepsilon _i^{(XT)}}$ and ${\rm \varepsilon _j^{(ET)}}$ are the site energies for the XT states $|e_i\rangle$ and the ET states $|c_j\rangle$, $J_{ij}$ is the XT coupling between $|e_i\rangle$ and $|e_j\rangle$, and $t_e$ is the ET coupling. The total Hamiltonian is then given by 
\begin{eqnarray}
 {\cal \hat H} &=& {\cal \hat H_S} - {\sum_{k=1}^{N_B}} {{{\hat V}_k}\sum\limits_{{\alpha_k}} {{g_{{\alpha_k}}}{{\hat x}_{{\alpha_k}}}} }  \nonumber \\
&+& {\sum_{k=1}^{N_B}} \sum\limits_{{\alpha_k}} {\left(\frac{{\hat p_{{\alpha_k}}^2}}{{2{m_{{\alpha_k}}}}} + \frac{1}{2}{m_{{\alpha_k}}}\omega _{{\alpha_k}}^2\hat x_{{\alpha_k}}^2\right)},
\end{eqnarray}
where ${\hat x}_{\alpha_k}$, ${\hat p}_{\alpha_k}$, $m_{\alpha_k}$, and ${\omega}_{\alpha_k}$ are the coordinate, momentum, mass, and frequency of the $\alpha_k$th oscillator for the $k$th bath, ${\hat V_k}$ is the system part of the couplings for the XT and ET states, {and $N_B$ is the number of the heat-bath.} Using this model, we investigated the efficiency of XCET process by changing {the number of sites ($N=4-6$), the energy configuration of the XT and ET states, the number of baths ($N_B=1-7$), and the form of $V_k$ as well as the coupling strength of the system-bath interactions.} Then we found that the system with weak off-diagonal system-bath interaction between the XT and ET bridge sites is essential
in order to realize efficient irreversible XCET process, due to the suppression of the quantum coherence between the XT and ET processes. {Here we show two representative results for the models that are inspired by photosynthetic systems.}

While the photosynthesis antenna system consists of more than ten XT states, in the present study, we set $n_{XT}=4$ and $N=6$, due to a limitation of CPU power (see Fig. \ref{fig:model}). Thus, $|e_4\rangle$ is set to be the initial state of the ET process, while $|c_5\rangle$ and $|c_6\rangle$ are charge transfer states. 
We considered systems with the six baths described by the diagonal interaction, $\hat V_{k} \equiv |{e_k}\rangle\langle {e_k}|$ for $k = 1-6$, for the XT and ET states, and one independent bath (XCET bath) described by the off-diagonal interaction, $\hat V_{7} \equiv |{e_{3}}\rangle\langle {e_{4}}|+|{e_{4}}\rangle\langle {e_{3}}|$, for the XCET process. {While the off-diagonal interactions between each site are weak,\cite{Renger2012} we found that in the XCET process cannot be neglected even very weak.} We assume that the XT and XCET states, $|{e_k}\rangle$ are coupled to overdamped Drude baths (the XT and XCET baths) described by the spectral distribution $J_k(\omega ) = {2 \lambda_k \gamma_k \omega }/ ({\omega ^2} + {\gamma_k ^2})$ \cite{Tanimura88,Ishizaki05,Tanimura06,Tanimura14,Tanimura15} for {$k=1-4$ and $7$,}
while the ET states, $|c_k\rangle$  are coupled to a Brownian mode bath (the ET bath) described by $ J_k(\omega)= {2 \lambda_k\gamma_k \omega_0^2 \omega }/[{(\omega_0^2-\omega^2)^2+\gamma_k^2\omega^2}]$\cite{Arend2015,TMJPSJ1994,TanakaJPSJ09,TanakaJCP10,TanimruaJCP12} for {$k$= 5 and 6}.
In the case of photosynthesis, the XT bath consists of the collective modes of the antenna system coupled to the XT states, whereas the ET bath consists of the collective modes of the reaction center coupled to the ET states. While the bath coupling strengths for the XT and ET baths are not weak, that for the XCET bath, which consists of the collective modes between the chlorophyll molecules of the antenna and the reaction center, is assumed to be weak, because the distance between the XT and ET systems is assumed to be large. For the same reason, we chose $J_{34}$ to be small. An off-diagonal XCET bath coupling may arise from the stretching mode between the XT and ET bridge sites. 

In order to analyze the role of the baths in the efficiency of the XCET process, we investigated the time evolution of the reduced density matrix using the HEOM approach.\cite{Arend2015,TMJPSJ1994,TanakaJPSJ09,TanakaJCP10,TanimruaJCP12}
With this approach, we can investigate system-bath interactions under non-perturbative, non-Markovian conditions. 
The bath is characterized by the noise correlation function, $C_k(t) \equiv \langle \hat{X}_k(t) \hat{X}_k(0) \rangle_\mathrm{B}$, where $\hat{X}_k \equiv \sum_j g_{\alpha_k}x_{\alpha_k}$ is the collective bath coordinate of the $k$th bath and $\langle \ldots \rangle_\mathrm{B}$ represents the average taken with respect to the canonical density operator of the baths. In the case that the noise correlation function, $C_k(t)$, takes the form of a linear combination of exponential functions and a delta function, as $C_k(t) = \sum_{j_k =0}^{J_k} ( c'_{j_k} + i c''_{j_k}) e^{-\gamma_{j_k}|t|} + 2\Delta_k \delta(t)$, which is realized in the cases of Drude \cite{Tanimura88,Ishizaki05,Tanimura06,Tanimura14, Tanimura15} and Brownian\cite{Arend2015,TMJPSJ1994,TanakaJPSJ09,TanakaJCP10,TanimruaJCP12} baths, 
we can derive the HEOM that consist of the following set of equations of motion for the auxiliary density operators (ADOs):
\begin{align}
&\frac{\partial}{\partial t} \hat{\rho}_{\bold{n}_1, \ldots, {\bold{n}_{N_B}}}(t) \nonumber\\
&=  - \left[ \frac{i}{\hbar} \hat {\mathcal{L}}
      + {\sum_{k=1}^{N_B}} \sum_{j_k =0}^{J_k} n_{j_k} \gamma_{j_k} \right]
      \hat{\rho}_{\bold{n}_1, \ldots, {\bold{n}_{N_B}}}(t)
\notag \\
   &\quad - {\sum_{k=1}^{N_B}} \Delta_k \hat{\Phi}_k^2
      \hat{\rho}_{\bold{n}_1, \ldots, {\bold{n}_{ N_B }}}(t)
\notag \\
   &\quad - {\sum_{k=1}^{N_B}} \hat{\Phi}_k \sum_{j_k=0}^{J_k}
      \hat{\rho}_{\ldots, \bold{n}_k + \bold{e}_{j_k},\ldots} (t)
\notag \\
  &\quad - {\sum_{k=1}^{N_B}} \sum_{j_k=0}^{J_k} n_{j_k} \hat{\Theta}_{j_k}
      \hat{\rho}_{\ldots, \bold{n}_k - \bold{e}_{j_k}, \ldots}(t).
\label{eq:HEOM}
\end{align}
Here $\bold{e}_{j_k}$ is the unit vector along the ${j_k}$th direction, and we have defined $\hat {\mathcal{L}} \hat{\rho} \equiv [ \hat{H}_\mathrm{S}, \hat{\rho} ]$, $\hat{\Phi}_k \hat{\rho} \equiv (i/\hbar) [\hat{V}_k, \hat{\rho} ]$, and $\hat{\Theta}_{j_k} \equiv c'_{j_k} \hat{\Phi}_k - c''_{j_k} \hat{\Psi}_k$ with $\hat{\Psi}_k \hat{\rho} \equiv (i/\hbar) \{\hat{V}_k, \hat{\rho} \}$.
Each ADO is specified by the index $\bold{n}_k = ( n_{k_{0_k}}, \ldots, n_{k_{J_k}})$, with {$N_B = 7$},
where each element takes an integer value greater than zero.
The ADO for which all elements are zero, $\bold{n}_1 =  \ldots  = {\bold{n}_{N_B}} = 0$, corresponds to the actual reduced density operator.

{As the initial conditions, we set the populations to $\langle e_1 | \hat \rho_{\bold {0, \ldots, 0}} | e_1 \rangle$=$1$ and the other diagonal elements of the reduced density matrix to 0 for simplicity. Note that we also tested the Boltzmann equilibrium states for the XT system as the initial condition, but the efficiency of the XCET process was same as in the present case besides initial temporal behaviors.} We fixed the characteristic frequency of the fifth bath in the up-and-down case to ${\rm {\omega}_0 = 500 cm^{-1}}$ and used this as the frequency unit for the system. Throughout the present investigation, we fixed the inverse temperature as $\rm {\beta}{\hbar}{\omega}_0 = 2.4 $(300 K).

{To demonstrate a role of the off-diagonal XCET bath interaction in a generalized manner, the site energies of the XT and ET states were chosen to be (i) the up-and-down and (ii) downhill configurations that were inspired by photosynthesis systems\cite{Nakatsuji1998, Yan2017, Mukamel2013} with adapting the parameters in the order of magnitudes as those of the PS II.\cite{Renger2017,Valkunas2017}} 
We constructed an energy scheme for the XT sites by adapting the funnel concept, in which, as XT proceeds, the site energy of each state decreases.\cite{Blankenship} {Although the ET states of PSII consists of two sites,\cite{Renger2017,Valkunas2017} here we consider three site for the ET states to explore the effects of coherence among the ET sites.}
Then, in both cases, the site energies of the XT states were chosen as ${\rm \varepsilon _1^{(XT)}} = 0.6{\omega}_0$, ${\rm \varepsilon _2^{(XT)}} = 0.6{\omega}_0$
${\rm {\varepsilon _3^{(XT)}} = 0.2{\omega}_0}$, ${\rm {\varepsilon _4^{(XT)}} = 0.0}$, {while those of the ET states were chosen to be (i) ${\rm \varepsilon _5^{(ET)}} = {\omega}_0$ and ${\rm \varepsilon _6^{(ET)}} = 0.0$ for the up-and-down case and (ii) ${\rm \varepsilon _5^{(ET)}} = -0.6{\omega}_0$ and ${\rm \varepsilon _6^{(ET)}} = -1.2{\omega}_0$ for the downhill case. In the downhill case, the characteristic frequency of the fifth bath was chosen as $\omega_0'=0.6\omega_0$.} 
For the system coupling parameters, we set $J_{12}=0.5{\omega}_0$, $ J_{23}=0.5{\omega}_0$ $J_{34}=0.01{\omega}_0$, and $t_e = 0.1{\omega}_0$, which are appropriate for a photosynthetic system. We chose the XCET system coupling, $J_{34}$, to be small in comparison with $J_{12}$ and $J_{23}$, reflecting the fact that the distance between the chlorophyll molecules in the antenna complex and the reaction center is large. We set all other $J_{ij}$ to 0. The bath coupling strengths (reorganization energy) were chosen 
as ${\lambda}_k = 0.2{\omega}_0$ for {$k=1-3$, ${\lambda}_4 =0.1{\omega}_0$, and ${\lambda}_k = 2.5{\omega}_0$ for $k=5$ and $6$. For the seventh off-diagonal XCET bath coupling, we considered (a) the non-XCET (${\lambda}_7 =0$), (b) very weak (${\lambda}_7 =0.0001{\omega}_0$), (c) weak (${\lambda}_7 =0.001{\omega}_0$), and (d) moderate (${\lambda}_7 =0.01{\omega}_0$) cases, in comparison with $J_{34}=0.01$.} {The inverse noise correlation time were chosen as ${\gamma_k} = 0.1{\omega}_0$ for $k = 1-4$ and ${\gamma_7} = 0.5{\omega}_0$ with  (i) ${\gamma_k} = 1.0{\omega}_0$ for $k = 5$ and $6$ in the up-and-down case and with (ii) ${\gamma_k} = 0.6{\omega}_0$ for $k = 5$ and $6$ in the downhill case.}
The HEOM given in eq \eqref{eq:HEOM} were then numerically integrated using the fourth-order exponential integrator method, with a time step of 0.01 / ${\omega}_0$. We chose the truncation number of the hierarchy, representing the depth of the HEOM computation, as { ${\rm n}_{k_{max}} = 5 $ for the Drude baths, and ${{\rm n}_{k_{max}} = 8}$} for the Brownian baths. A Pad{\'e} spectral decomposition scheme was employed to obtain the expansion coefficients of the noise correlation functions. \cite{YanPade10A, Hu} {We set the maximum number of the hierarchy levels to be 18.}

\begin{figure}[tbp]
  \centering
  \includegraphics[clip, width = 6.0cm]{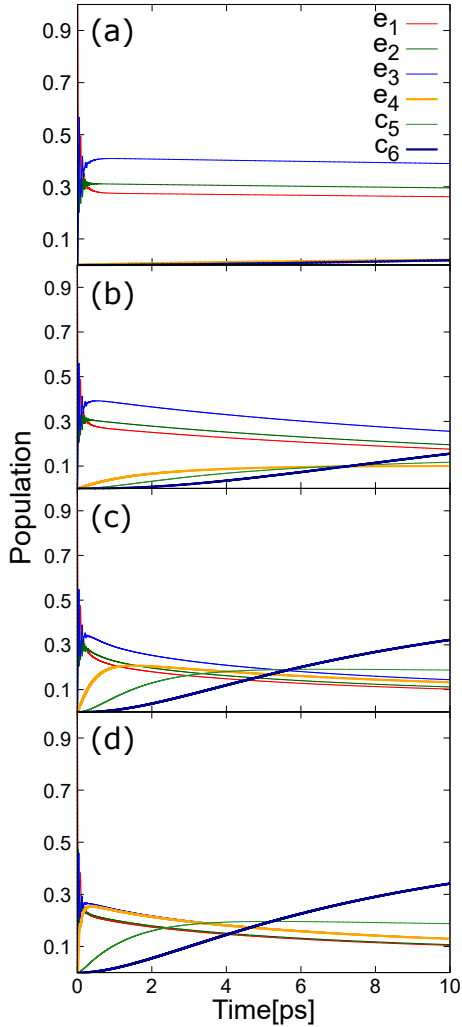}
  \caption{Time evolution of the density matrix {for (i) the  up-and-down model in (a) the non-XCET, (b) very weak, (c) weak, and (d) moderate off-diagonal XCET bath coupling cases.} In this figure, the solid curves represent the populations of the XT and ET states, respectively.} 
  \label{fig:up-and-down}
\end{figure}

\begin{figure}[tbp]
  \centering
  \includegraphics[clip, width = 6.0cm]{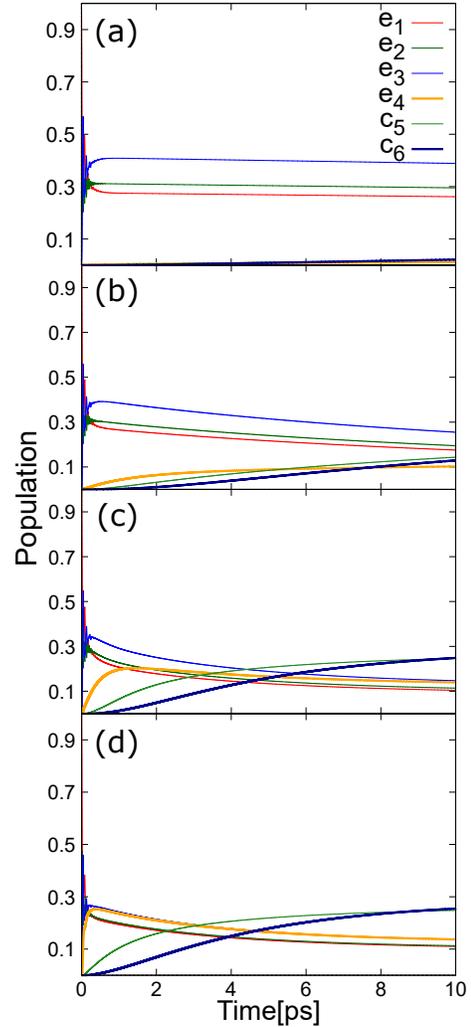}
  \caption{Time evolution of the density matrix {for (ii) the  downhill model in (a) the non-XCET, (b) very weak, (c) weak, and (d) moderate off-diagonal XCET bath coupling cases,} In this figure, the solid curves represent the populations of the XT and ET states, respectively.}
  \label{fig:downhill}
\end{figure}

{Figures \ref{fig:up-and-down} and \ref{fig:downhill} illustrate the time evolution of the density matrix elements for (i) the up-and-down and (ii) downhill models for the different off-diagonal XCET bath couplings. In all cases, besides the initial temporal oscillations, the $|e_4\rangle$ population increase following the decrease of the population of XT states. Consequently, the $|c_6\rangle$ population increases through the $J_{34}$ coupling. }

{As illustrated in Figs. \ref{fig:up-and-down}(a)-\ref{fig:up-and-down}(d) and \ref{fig:downhill}(a)-\ref{fig:downhill}(d), the efficiency of the XCET transition is very low without the off-diagonal XCET bath interaction.} This is because pure dephasing is the dominant effect in the diagonal bath coupling { (i.e. $V_k$ with ${\gamma_k} = 0.1{\omega}_0$ for $k=1-4$) for the present slow modulation case,} while population relaxation is the dominant effect in the off-diagonal bath coupling.\cite{IshizakiJCP2006} For this reason, because pure dephasing merely causes fluctuations in the energies of exciton at individual sites, and does not contribute to the transported energy, the XT from the $|e_3\rangle$ site to the $|e_4\rangle$ site is small. {Thus, the population of $|c_6\rangle$ is also small.}

{As illustrated in Figs. \ref{fig:up-and-down}(b) and \ref{fig:downhill}(b), the efficiency of the XCET process is dramatically enhanced under the presence of the off-diagonal XCET bath interaction, even it is very weak.} This is because coherence in the XT process is suppressed by the XCET bath in the bridge site, $|e_4\rangle$, due to the off-diagonal XCET bath interaction that causes the population relaxation from the $|e_3\rangle$ site to the $|e_4\rangle$ site. {Although the energy configurations are very different, the dynamical aspects of up-and-down and downhill cases are similar. This indicates that the transition from the $|e_4\rangle$ site to the $|c_6\rangle$ site is not a thermal origin but a quantum origin that arises from the $t_e$ interactions. After the $|e_4\rangle$ site is populated, a coherent ET transition occurs from the $|e_4\rangle$ site to the $|c_6\rangle$ site via the $|c_5\rangle$ site.\cite{Sumi96,TanakaJPSJ09,TanakaJCP10,TanimruaJCP12} }

{As illustrated in Figs. \ref{fig:up-and-down}(b)-\ref{fig:up-and-down}(d) and Figs. \ref{fig:downhill}(b)-\ref{fig:downhill}(d), the increase of the $|c_6\rangle$ population does not follow the quick increase of the $|e_4\rangle$ population.} This is because the ET transition is slow in comparison with the $|e_3\rangle$-$|e_4\rangle$ transition so that the ET process become the rate-determine step of the XCET process. The equilibration time of the $|e_4\rangle$ site that we obtain from our results in the weak coupling case is estimated as 30ps, which is consistent with the experimentally determined time scale for the XT process taking place between the antenna complex and the reaction center in a photosynthetic system.

After the population reaches the $|c_6\rangle$ site, it does not return to the $|e_4\rangle$ site, because, {due to the influence of thermal effect, at such a time, the distribution goes to the lower energy sites.} While the steady state population of $|e_4\rangle$ increases from zero to the weak XCET bath coupling region, as illustrated in Figs. \ref{fig:up-and-down}(a)-\ref{fig:up-and-down}(c) and Figs. \ref{fig:downhill}(a)-\ref{fig:downhill}(c), the increase of the populations is suppressed in the moderate coupling region, as illustrated in Figs. \ref{fig:up-and-down}(d) and \ref{fig:downhill}(d).  {This is because the efficiency of off-diagonal XCET transition has already reached the maximum in the weak coupling region.} As the result, the population of the $|c_6\rangle$ state does not increase as the increase of the XCET coupling strength above the moderate coupling region.

These results indicate that the off-diagonal XCET bath coupling plays an essential role in the XT-ET conversion process. Even in the weak exciton coupling regime ($J_{34} = 0.01$), the enhancement is particularly strong for the $|e_4\rangle$ site, which for a photosynthetic system consists of the transfer of excitation energy from the antenna complex to the reaction center. While a strong XCET bath coupling may not be realistic for photosynthetic systems,\cite{Renger2012} a mechanism utilizing an off-diagonal weak coupling bath is conceivable for realizing efficient XT-ET conversion. 

The above results indicate that to realize efficient XT-ET conversion, coherence in the XT process must be suppressed in the bridge site, $|e_4\rangle$, because otherwise the population in this site does not increase, due to the collective coherent motion in the XT sites. This implies that the XT-ET conversion occurs consecutively rather than concertedly, due to the interaction with the XCET bath. This mechanism does indeed enhance the conversion rate.  

{Although the present investigations are limited to the specific models, we believe that the applicability of our finding for the off-diagonal XCET bath is wider.} In future investigations, we plan to extend the present model to employ more realistic parameter values {for photosynthetic system\cite{Nakatsuji1998,Yan2017,Mukamel2013, Renger2005, Renger2012,Renger2015, Renger2017, Valkunas2017} and organic photovoltaic materials\cite{Tamura2013, Prior2017} to investigate a role of the off-diagonal XCET bath interaction.} 
\section{Acknowledgments}
This research is supported by a Grant-in-Aid for Scientific Research (A26248005) from the Japan Society for the Promotion of Science.

\end{document}